\documentclass[aps,pre,twocolumn,superscriptaddress]{revtex4-1}
\usepackage[dvips]{graphicx}
\begin{document}

%\preprint{}

\title{Balance network of asymmetric simple exclusion process}


\author{Takahiro Ezaki}
\email{ezaki@jamology.rcast.u-tokyo.ac.jp}
%\thanks{}
%\altaffiliation{dept}
\affiliation{Department of Aeronautics and Astronautics, School of Engineering,
The University of Tokyo, 7-3-1, Hongo, Bunkyo-ku, Tokyo 113-8656, Japan}
\author{Katsuhiro Nishinari}
\email{tknishi@mail.ecc.u-tokyo.ac.jp}
%\thanks{}
%\altaffiliation{rcast}
\affiliation{Research Center for Advanced Science and Technology,
The University of Tokyo, 4-6-1, Komaba, Meguro-ku, Tokyo 153-8904, Japan}


\date{\today}

\begin{abstract}
We investigate a balance network of the asymmetric simple exclusion process (ASEP). Subsystems consisting of ASEPs are 
connected by bidirectional links with each other, which results in balance between every pair of subsystems. The network includes some 
specific important cases discussed in earlier works such as the ASEP with the Langmuir kinetics, multiple lanes and finite reservoirs. 
Probability distributions of particles in the steady state are exactly given in factorized forms according to their balance properties.
Although the system has nonequilibrium parts, the expressions are well described in a framework of statistical mechanics based on equilibrium states.
Moreover, the overall argument does not depend on the network structures, and the knowledge obtained in this work is applicable to a broad range of problems. 
\end{abstract}

%\keywords{}

\maketitle
\section{introduction}
The asymmetric simple exclusion process (TASEP) is one of the most paradigmatic models to understand phenomena in nonequilibrium physics \cite{NE}.
The model, consisting of a one-dimensional lattice and particles with hard-core exclusion interaction, describes fundamental transport phenomena and is
applied to a broad range of problems: traffic flow \cite{vt}, biological transport \cite{lk,Kin,Kin1,Kin2}, etc.
As natural extensions of the TASEP, the effects of particle attachments and detachments in the bulk \cite{lk,lk1,lk2}, and multiple lanes \cite{ml,ml2,ml3,ml4,ml5,ml6,ml7} have been investigated, 
and some significant results have been presented. 
These systems allow additional motion of particles in the TASEPs and can be interpreted as networks of TASEPs and reservoirs, 
where each site in a lattice is connected with the particle reservoir or a site in a different TASEP.
On the other hand, the TASEP on networks has been focused on recently \cite{net,net1,net2,net3}. The results have concluded that the dynamics of
the system depends on structure of the networks. In this paper, we focus on an exactly solvable 
network consisting of the periodic TASEPs. The steady state of the system is described by general expressions, which are found to be \textit{independent} of 
the topology of the network. The key to construct the expressions is detailed balance satisfied among the subsystems; in other words, 
the TASEPs are in balance with each other in the network. 
This kind of structure has been reported in the previous studies \cite{lk2,ml}, and we have successfully generalized the system in this work.
We provide a certain class of solvable TASEP systems, which includes some important models.

The rest of this paper is organized as follows. Section \ref{model} gives a definition of the network and the relationship to relevant models. 
In Sec. \ref{exact}, we give the exact stationary distribution of the system. Using these expressions, we derive some physical quantities in Sec. \ref{cor}.
Finally, we summarize the discussion in Sec. \ref{con}.

\begin{figure}[htbp]
 \begin{center}
  \includegraphics[width=80mm]{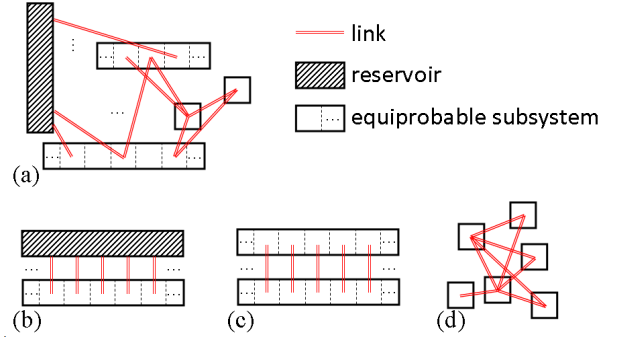}
 \end{center}
 \caption{Examples of the balance network. The balance network can be regarded as generalization of (b) the TASEP with the Langmuir kinetics, (c)
 the multilane TASEP, and (d) the simple exclusion network.
 }
 \label{net}
\end{figure} 
\section{Model}\label{model}
We consider a network of exclusion process consisting of particles, sites, links and a single reservoir. 
Each site can contain at most one particle, and each particle jumps to a site in the same subsystem or 
to a site in another subsystem through a link. Here the TASEP on a
ring is mainly focused on as the subsystem. The periodic TASEP has an `equiprobable' property that all the configurations of particles,
$\{\tau_i^j\}_j$, appear equally likely in the steady state,  when it is isolated.

 Here, $\tau_i^j$ is the occupation 
number of site $i$ ($i=1,\cdots,L_j$) in the subsystem $j$ ($j=1,\cdots,K$), and $\{\tau_i^j\}_j$ is a set of the occupation numbers that describes each configuration
in subsystem $j$.
In principle, other exclusion processes (or even processes of bosons discussed later) 
can also be candidates for this subsystem if only they satisfy the equiprobable property.
In this work, we also consider a single site without dynamics in itself as one of the equiprobable subsystems. 
The components in the system are summarized as follows:

(i) Equiprobable subsystem:
A set of sites that has equiprobable dynamics such as the TASEP with periodic boundary conditions.
Each site in an equiprobable system $j$ has a common leaving rate of particles, $\chi_j$; 

(ii) Link:
Bidirectional links connect pairs of sites in different subsystems (unidirectional links are forbidden). 
Each site can have an arbitrary number of links;

(iii) Reservoir:
A reservoir can accept and provide an arbitrary number of particles through links, and its provision rate is $\chi_R$. 
Only a single reservoir is allowed in the system.

\begin{figure}[htbp]
 \begin{center}
  \includegraphics[width=70mm]{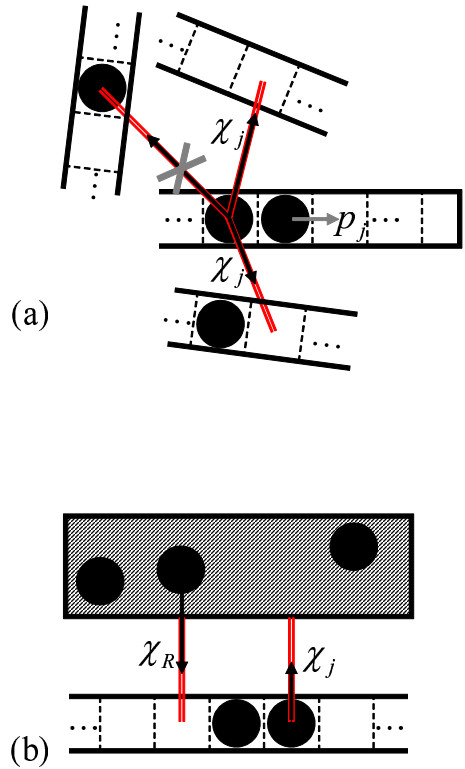}
 \end{center}
 \caption{Transition rules of the system. 
 Particles hop to the next site with a rate $p_j$ (the TASEP) and leave the subsystem $j$ with a rate $\chi_j$ through the links,
  obeying the exclusion principle. The reservoir contains an infinite number of particles and
  provides a particle to an empty linked site with a rate $\chi_R$.}
 \label{rule}
\end{figure}

These components are set in the system as a network [see Fig. \ref{net} (a)].
Note that the network must be a \textit{connected network}: the network cannot have isolated parts, and every pair of subsystems must be interconnected by links and/or other subsystems.
In the network particles jump to the neighboring sites, following the hard-core exclusion principle.
As shown in Fig. \ref{rule} (a), a particle at a randomly chosen site in the subsystem $j$ jumps to the next site to its right with a rate $p_j$, and to linked sites 
in other subsystems with a rate $\chi_j$, if the target sites are empty. Moreover, through the links,
the reservoir can accept and provide particles with rates $\chi_j$ and $\chi_R$, respectively [see Fig. \ref{rule} (b)].

By these formulations of the system, we can see that the network includes some important cases; the TASEP with the Langmuir kinetics \cite{lk} [Fig. \ref{net} (b)],
the multilane TASEP \cite{ml,ml2} [Fig. \ref{net} (c)], and a simple exclusion network [Fig. \ref{net} (d)].
Furthermore, the balance network generally represents multiple competing TASEPs.
In the context of biology, the competition of the TASEPs is discussed as a problem of multiple mRNAs \cite{competition}, or it may explain the dynamics of motor proteins
on a spindle consisting of microtubules in cell division.

%%%%%%%%%%%%%%%%%%%%%%%%%%%%%%%%%%%%%%%%%%%%%%%%%%%%%%%%%%%%%%%%%%%%%%%%%%%%%%%%%%%%%%%%%%%%%%%%%%%%%%%%%%%%%%%%%%%%%%%%%%%%%%%%%%%%
\section{Exact analyses}\label{exact}
We analyze the balance network in the steady state, focusing on the probability distribution for each configuration of particles.
First let us review the expressions for the TASEP on a ring. A possible configuration $\{\tau_i\}$ is realized with the probability 
\begin{equation}
P(\{\tau_i\}) = N^{-1} f (\{\tau_i\}),
\end{equation}
where $f(\{\tau_i\})$ is the probability weight of each configuration, and $N^{-1}$ is the normalization factor.
Since all the possible configurations in this system are equally likely in this system, 
$f(\{\tau_i\})=1$. This property does not depend on the system length, $L_j$, and the density of particles.

Then, we present the probability distributions for the balance network using the weight of each configuration in subsystem $j$ with $n_j$ particles, $f_{n_j}(\{\tau_i^j\}_j)=1$.
The probability of finding the system in a configuration $(\{\tau_{i}^1\}_1,\cdots,\{\tau_{i}^K\}_K)$ is given by
\begin{eqnarray}
P(\{\tau_{i}^1\}_1,\cdots,\{\tau_{i}^K\}_K)&=&\Xi^{-1}\prod_{j=1}^{K}\left(\frac{\chi_R}{\chi_j}\right)^{n_j}f_{n_j}(\{\tau_i^j\}_j)\nonumber\\
&=&\Xi^{-1}\prod_{j=1}^{K}\left(\frac{\chi_R}{\chi_j}\right)^{n_j},\label{gc}
\end{eqnarray}
\begin{equation}
\Xi = \sum^{L_1}_{n_1=0}\cdots\sum^{L_K}_{n_K=0}\prod_{j=1}^{K}{\left(\frac{\chi_R}{\chi_j}\right)^{n_{j}}}\left(
\begin{array}{c}
L_j  \\
n_j 
\end{array}
\right).\label{Xi}
\end{equation}
Here $\Xi^{-1}$ is the normalization factor. As shown in the next part, this $\Xi$ corresponds to the grand partition function in statistical mechanics.
Note that, even if the system lacks the reservoir, this expression can be used with a slight modification.
In this case, since the absence of the reservoir leads to a constraint on the particle number, the sum of the weights is taken over all the configurations with a
given number of particles, $n$, while $\Xi$ is obtained by considering all the possible configurations for any particle numbers. 
Furthermore, the provision rate $\chi_R $ does not influence the equations because each $\chi_R^{n}$ in Eq. (\ref{gc}) is cancelled out by the normalization factor.
Although the conservative systems are also interesting when we consider actual biological processes with finite resource \cite{ml,competition,lim,lim2}, 
constraints on particle numbers often cause computational difficulty (see Ref. \cite{ml} for example). 

Let us confirm that these expressions correctly describe the system in the steady state by considering the master equation:
\begin{equation}
0=\frac{\partial }{\partial t}P(\mathcal{C}) = \sum_{\mathcal{C'}\neq \mathcal{C}}{\left\{P(\mathcal{C'})W(\mathcal{C'}\rightarrow \mathcal{C})-P(\mathcal{C})W(\mathcal{C}\rightarrow \mathcal{C'})\right\}}\label{mas},
\end{equation}
where $\mathcal{C}$ and $W$($\mathcal{C}\rightarrow\mathcal{C'}$) indicates the configuration of particles and the transition probability from configuration $\mathcal{C}$ to $\mathcal{C'}$, respectively. 
Here we separate the transitions into three parts, i.e., internal transitions in each subsystem, intersubsystem transitions, and transitions between the reservoir and subsystems.
Since each internal transition does not change the particle numbers in the subsystems, it is obvious that Eq. (\ref{gc}) satisfies the master equation for these transitions (for each subsystem, 
the equiprobable expression for a fixed particle number satisfies the master equation for the equiprobable subsystem, and thus the internal transition terms vanish). 
Generally, these terms vanish through taking the sum of all the transitions, and the detailed balance conditions are not satisfied: this is the generalization of the detailed balance to the nonequilibrium steady state \cite{ness}.
On the other hand, the other two types of transitions satisfy the detailed balance conditions:
\begin{equation}
0=P(\mathcal{C'})W(\mathcal{C'}\rightarrow \mathcal{C})-P(\mathcal{C})W(\mathcal{C}\rightarrow \mathcal{C'})\label{db}.
\end{equation}
Let us take a transition between subsystem $j_1$ and $j_2$ ($j_1<j_2$) as an example. In the transition, a particle jumps from site $i_1$ in subsystem $j_1$ to site $i_2$ in the subsystem $j_2$ through a link,
 which results in a change of the particle numbers in the subsystems, $\{\cdots,n_{j_1},\cdots,n_{j_2},\cdots\}\rightarrow \{\cdots,n_{j_1}-1,\cdots,n_{j_2}+1,\cdots\}$. Taking its reverse transition into account, 
Eq. (\ref{db}) holds:
\begin{eqnarray}
&&\Xi^{-1}\prod_{j=1,j\neq j_1,j_2}^{K}\left(\frac{\chi_R}{\chi_j}\right)^{n_j}\left[\left(\frac{\chi_R}{\chi_{j_1}}\right)^{n_{j_1}}\left(\frac{\chi_R}{\chi_{j_2}}\right)^{n_{j_2}}\chi_{j_1} \right.\nonumber\\
&&\qquad\qquad\qquad\qquad\quad\qquad -\left.\left(\frac{\chi_R}{\chi_{j_1}}\right)^{n_{j_1}-1}\left(\frac{\chi_R}{\chi_{j_2}}\right)^{n_{j_2}+1}\chi_{j_2}\right]\nonumber\\
&&=0.
\end{eqnarray}
Each bidirectional link ensures the existence of the reverse transition for a given intersubsystem transition. 
In the same manner, we can prove these detailed balance conditions for the transitions between the equiprobable subsystem and the reservoir.
Moreover, these cancellation mechanisms are independent of the network structure and capacity of sites; the expressions are valid for finite pools of particles \cite{pool}, 
and there is scope for extension of the subsystems to multiple occupation processes.

To summarize, all the terms in Eq. (\ref{mas}) vanish according to the properties of transitions, i.e., the nonequilibrium in the internal transitions
and the balance in the external transitions.

For the case of the TASEP network,
from Eqs. (\ref{gc}) and (\ref{Xi}) we can derive the current of particles in subsystem $j$ defined as $J_j = \langle \tau_i^j(1-\tau_{i+1}^j)\rangle$ by
considering all the configurations with $\tau_i^j=1$ and $\tau_{i+1}^j=0$:
\begin{eqnarray}
J_j &=& \Xi^{-1} \prod_{j'\neq j}^{K} \sum^{L_{j'}}_{n_{j'}=0}{\left(\frac{\chi_R}{\chi_{j'}}\right)^{n_{j'}}}\left(
\begin{array}{c}
L_{j'}  \\
n_{j'} 
\end{array}
\right)\nonumber\\
&&\qquad\qquad \times\sum^{L_{j}}_{n_{j}=0}{\left(\frac{\chi_R}{\chi_{j}}\right)^{n_{j}}}\left(
\begin{array}{c}
L_{j}-2  \\
n_{j}-1 
\end{array}
\right)\\
&=&\frac{\chi_j/\chi_R}{(\chi_j/\chi_R+1)^2}.\label{current}
\end{eqnarray}
In the first expression we swapped the summation and the multiplication. The current is the quantity of great importance
to describe transportation phenomena and the characteristic quantity in the nonequilibrium systems. 
It is noteworthy that the current is determined only by the parameters of the subsystem and the reservoir.
Moreover, we can prove that the correlation between the occupation numbers of successive two sites can be ignored even for finite size of the systems \cite{ind}. 

%%%%%%%%%%%%%%%%%%%%%%%%%%%%%%%%%%%%%%%%%%%%%%%%%%%%%%%%%%%%%%%%%%%%%%%%%%%%%%%%%%%%%%%%%%%%%%%%%%%%%%%%%%%%%%%%%%%%%%%%%%%%%%%%%%%%%%
\section{Correspondence to statistical mechanics}\label{cor}
The structure of the Eq. (\ref{gc}) is well explained in a framework of statistical mechanics.
Let us derive the expected value of the occupation numbers, putting $\chi_j = e^{\beta \epsilon_j}$ and $\chi_R = e^{\beta \mu}$ to emphasize the correspondence.
The grand partition function is calculated as
\begin{eqnarray}
\Xi(\beta,\mu)&=& \sum^{L_1}_{n_1=0}\cdots\sum^{L_K}_{n_K=0}\prod_{j=1}^{K}{e^{-\beta (\epsilon_j - \mu)n_j}\left(
\begin{array}{c}
L_j  \\
n_j 
\end{array}
\right)
}\\
&=& \quad\: \: \: \prod_{j=1}^{K}\sum_{n_j=0}^{L_k}{e^{-\beta (\epsilon_j - \mu)n_j}\left(
\begin{array}{c}
L_j  \\
n_j 
\end{array}
\right)
}\\
&=& \quad\: \: \:\prod_{j=1}^{K}\left(1+e^{-\beta(\epsilon_j - \mu)}\right)^{L_j}\\
&=& \quad\: \: \:\prod_{j=1}^{K}\Xi_{j}(\beta,\mu)\label{xik},
\end{eqnarray}
where $\Xi_{j}$ is defined as $\Xi_{j}=(1+e^{-\beta(\epsilon_j-\mu)})^{L_j}$.
Then, the expected value of the occupation number, $\langle n_j \rangle$ is given by
\begin{eqnarray}
\langle n_j \rangle &=& \frac{1}{\Xi(\beta,\mu)}\sum^{L_1}_{n_1=0}\cdots\sum^{L_K}_{n_K=0}n_j\prod_{j'=1}^{K}{e^{-\beta (\epsilon_{j'} - \mu)n_{j'}}}\\
&=&\frac{1}{\Xi_{j}(\beta,\mu)}\sum_{n_j=0}^{L_j}n_j {e^{-\beta (\epsilon_j - \mu)n_j}}\\
&=&\frac{1}{\beta}\frac{\partial }{\partial \mu}\log{\Xi_{j}(\beta,\mu)}\\
&=&\frac{L_j}{e^{\beta (\epsilon_j - \mu)}+1}\label{fermi}\\
&=&\frac{L_j}{\chi_j/\chi_{R} +1}.
\end{eqnarray}
Thus, the density of particles in each subsystem is derived. Note that these calculations can be performed without the interpretation using the energy, the inverse temperature, and the chemical potential;
however, the expressions are highly suggestive. If one regards each site as a distinctive energy state with energy, $\epsilon_j$, of fermions, Eq. (\ref{fermi}) coincides with the Fermi distribution 
[let $L_j=1$ for the simplicity of the argument ($L_j$ corresponds to the \textit{degeneracy})]. On the other hand, the system can also be interpreted as a problem of chemical adsorption with chemical potential, $\mu$, and stabilization energy, $-\epsilon_j$.
In this case, Eq. (\ref{fermi}) corresponds to the Langmuir isotherm of the system with independent $L_j$ sites in contact with the reservoir.
Since each pair of connected subsystems are in balance, the network is equivalent to a set of separated subsystems in balance with the reservoir.
Thus, the steady state is determined only by the parameters of each subsystem and the common reservoir. 
This is the reason why the overall argument can be well understood in the framework of statistical mechanics. 
However, it is still noteworthy that the statistical mechanics expressions can be naturally extended to the system consisting of some nonequilibrium parts.

Additionally, we evaluate the variance of the particle number in the subsystem $j$, $Var[n_j]$, as a physical quantity characterizing the equilibrium steady state.
\begin{eqnarray}
Var[n_j] &=& \langle n_j^2 \rangle - \langle n_j \rangle^2\\
&=& \frac{1}{\beta^2}\frac{\partial^2 }{\partial \mu^2} \log{\Xi_j(\beta,\mu)}\\
&=&\frac{L_je^{\beta (\epsilon_j - \mu)}}{(e^{\beta (\epsilon_j - \mu)}+1)^2}\\
&=&\frac{L_j\chi_j/\chi_{R}}{(\chi_j/\chi_{R} +1)^2}. 
\end{eqnarray}
Interestingly, the expression is associated with the current Eq. (\ref{current}) as
\begin{equation}
Var[n_j] = L_j J_j.
\end{equation}
Thus, an intriguing relation is derived, where the representative quantities of equilibrium and nonequilibrium physics are linked together. \cite{der}

%%%%%%%%%%%%%%%%%%%%%%%%%%%%%%%%%%%%%%%%%%%%%%%%%%%%%%%%%%%%%%%%%%%%%%%%%%%%%%%%%%%%%%%%%%%%%%%%%%%%%%%%%%%%%%%%%%%%%%%%%%%%%%%%%%%%%%
\section{Conclusions}\label{con}
We have presented the balance network consisting of nonequilibrium subsystems, bidirectional links, and a single reservoir.
The network includes a wide variety of models relevant to previous works and is very useful in the meaning of application.
On the other hand, the network has structure of balance connections, which allows us analytical solutions.
From the probability distribution of particles we can calculate some physical quantities, 
and the overall argument can be well understood in the framework of established statistical physics. 

In the balance network, only bidirectional links are allowed because unidirectional links will cause the 
`flow' of particles between subsystems and violate the balance relations. 
Besides, if we allow a single site which can contain more than one particle, 
the site is equivalent to a finite pool of particles or a finite reservoir.
The balance network can contain an arbitrary number of finite reservoirs; on the other hand,
construction of exact probability distribution for the system with multiple \textit{infinite} reservoirs is not
straightforward. In future works, further analyses on the extension of the balance network and its relations with nonequilibrium physics are needed.

\bibliography{ref}

\begin{thebibliography}{00}
\bibitem{NE}B. Derrida and M. Evans, in Nonequilibrium Statistical Mechanics in One Dimension, edited by V. Privman (Cambridge University Press, Cambridge, England, 1997), Chap. 14, pp. 277.304.
\bibitem{vt}D. Chowdhury, L. Santen and A. Schadschneider, Phys. Rep. 329 (2000) 199.
\bibitem{lk}A. Parmeggiani, T. Franosch and E. Frey, Phys. Rev. Lett. 90, 086601 (2003).
\bibitem{lk1}A. Parmeggiani, T. Franosch, and E. Frey, Phys. Rev. E 70, 046101 (2004). 
\bibitem{Kin}K. Nishinari, Y. Okada, A. Schadschneider, and D. Chowdhury, Phys. Rev. Lett. 95, 118101 (2005).
\bibitem{Kin1}P. Greulich, A. Garai, K. Nishinari, A. Schadschneider, and D. Chowdhury, Phys. Rev. E 75, 041905 (2007).
\bibitem{Kin2}D. Chowdhury, A. Garai, and J. S. Wang, Phys. Rev. E 77, 050902(R) (2008).
\bibitem{lk2}T. Ezaki and K. Nishinari, J. Phys. A 45, 185002 (2012).
\bibitem{ml}T. Ezaki and K. Nishinari, Phys. Rev. E 84, 061141 (2011).
\bibitem{ml2}V. Popkov and I. Peschel, Phys. Rev. E 64, 026126 (2001).
\bibitem{ml3}H.-W. Lee, V. Popkov, and D. Kim, J. Phys. A 30, 8497 (1997).
\bibitem{ml4}R. Jiang, M.-B. Hu, Y.-H. Wu, and Q.-S. Wu, Phys. Rev. E 77, 041128 (2008).
\bibitem{ml5}E. Pronina and A. B. Kolomeisky, J. Phys. A 37, 9907 (2004).
\bibitem{ml6}R. J. Harris and R. B. Stinchcombe, Physica A 354, 582 (2005).
\bibitem{ml7}T. Mitsudo and H. Hayakawa, J. Phys. A 38, 3087 (2005).
\bibitem{net}I. Neri, N. Kern and A. Parmeggiani, Phys. Rev. Lett. 107, 068702 (2011).
\bibitem{net1}R. Stinchcombe, Physica A 346, 1 (2005).
\bibitem{net2}M. Basu, P.K.Mohanty, J. Stat. Mech., P10014 (2010).
\bibitem{net3}B. Embley, A. Parmeggiani, and N. Kern, Phys. Rev. E 80, 041128 (2009). 
\bibitem{competition}L. J. Cook, R. K. P. Zia, and B. Schmittmann, Phys. Rev. E 80, 031142 (2009).
\bibitem{lim}C. A. Brackley, M. C. Romano, C. Grebogi, and M. Thiel, Phys. Rev. Lett. 105, 078102 (2010).
\bibitem{lim2}P. Greulich, L. Ciandrini, R. J. Allen, and M. C. Romano, Phys. Rev. E 85, 011142 (2012).
\bibitem{ness}B. Derrida, J. Stat. Mech., P07023 (2007).
\bibitem{pool}A finite pool can contain a finite number of particles and cannot have distinct configurations of particles for a given particle number.
(Binomial factors in the partition function are replaced by one.)
\bibitem{ind}The density $\rho_j =\langle \tau_i^j \rangle$ is given as $\rho_j = \frac{1}{\chi_j/\chi_R+1}$ which leads to $J_j = \rho_j(1-\rho_j)$, in the same manner.
\bibitem{der}Corresponding to this equation,
the variance of the occupation number $Var[\tau_i^j]$ can also be expanded as $Var[\tau_i^j]=\langle \tau_i^{j2} \rangle-\langle \tau_i^{j} \rangle^2
=\langle \tau_i^{j} \rangle-\langle \tau_i^{j} \rangle^2=\rho_j-\rho_j^2=J_j$.
\end{thebibliography}

\end{document}